\documentstyle[12pt,aasms4]{article}

\begin{document}

\title{Neutrino oscillations and gamma-ray bursts}
\author{W. Klu\'zniak}
\affil{Copernicus Astronomical Center, Bartycka 18, 00-716 Warszawa,
Poland\\
and University  of Wisconsin, Sterling Hall,
1150 University Ave., Madison, WI 53705}
\authoremail{wlodek@astrog.physics.wisc.edu}

\begin{abstract}
If the ordinary neutrinos oscillate into a sterile flavor in a manner
consistent with the Super-Kamiokande data on the zenith-angle
dependence of atmospheric $\mu$-neutrino flux, an energy sufficient to
power a typical cosmic gamma-ray burst (GRB) ($\sim10^{52}$ erg) can be
carried by sterile neutrinos away from the source and deposited in a
region relatively free of baryons. Hence, ultra-relativistic bulk
motion (required by the theory of and observations of GRBs and their
afterglows) can easily be achieved in the vicinity of plausible
sources of GRBs.  Oscillations between sterile and ordinary neutrinos
would thus provide a solution to the ``baryon-loading problem'' in the
theory of GRBs.
\end{abstract}


\section{Ultra-relativistic bulk motion and the baryon-loading problem}

Available evidence clearly shows that GRBs originate in distant
reaches of the universe (Fishman \& Meegan 1995),
with redshifts $z>0.8$ and $z=3.4$
reported for two particular sources (Djorgovski {\it et al.} 1997,
Kulkarni {\it et al.} 1998).

Ultra-relativistic bulk motion of matter in
the sources is required to account for the observed fluences
($10^{-5 \pm 2}$erg/cm$^2$) in cosmic gamma-ray bursts originating at a
redshift of $z\sim 1$ (M\'esz\'aros \& Rees 1993).
Otherwise, if the source were sufficiently
large to be optically thin to electron pair creation the typical
duration of a GRB would be days, rather than the observed seconds or
minutes, while a source only light minutes (or less) in size would be
opaque to its own gamma-ray radiation. Considerations of such bulk
motion led to predictions (Paczy\'nski \& Rhoads 1993, Katz 1994,
M\'esz\'aros \& Rees 1997, Vietri 1997) of radio, optical and X-ray afterglows,
which were subsequently observed (Galama {\it et al.} 1997, van Paradijs
{\it et al.} 1997, Frail {\it et al.} 1997).

However, most sources capable of impulsively releasing the $10^{52}$
erg or more of energy required to power a GRB contain so much matter
around them, that if the energy released were used to accelerate even a
very small fraction ($\sim 10^{-3}$) of the baryons present, only
a non-relativistic wind would result. This is known as the baryon-loading
problem (e.g. Piran 1997).

It has been hoped that the geometry of the sources is such that at
least some of the energy released is channeled along directions
relatively free of baryons, so that relativistic bulk motion and the
ensuing beaming of radiation may occur along certain lines of
sight. So far this has not yet been fully demonstrated for any
theoretical source of GRBs---although calculations of binary
``mergers'' of a neutron star with another neutron star or a black
hole suggest that this may be the case (Rasio \& Shapiro 1992;
Klu\'zniak \& Lee 1998; Wilson, Salmonson \& Mathews 1998), for other
models not even approximate calculations of the spatial distribution
of baryons is available (Woosley 1993, Paczy\'nski 1998).

The diversity of the observed GRBs is such, that it is not at all clear
that they all arise in sources of identical, or even similar,
geometry. One is tempted
to search for a solution to the baryon loading problem, which does not
invoke a very special geometry. The transport of energy by non-interacting
particles would provide a natural solution. Sterile neutrinos, if they exist,
would be such particles.

\section{The Super-Kamiokande result}

The Super-Kamiokande collaboration has found strong evidence for a zenith angle
dependence of the flux ratio of atmospheric
muon and electron neutrinos, $\nu_\mu/\nu_e$,
in a 535 day (33.0 kiloton-year) observation of \`Cerenkov photons in
a water detector (Fukuda {\it et al.} 1998).
Oscillation between the muon and the electron neutrino is not
favored by the data. The data is well fit by
a two-neutrino model of vacuum oscillations of the $\mu$ neutrino
``$\nu_\mu \leftarrow\rightarrow \nu_x$, where $\nu_x$ may be $\nu_\tau$ or
a new,
non-interacting `sterile' neutrino'' ({\it ibid.}). The probability for
a flavor change in travel over a distance $L$ is given by
$$P_{a\rightarrow b}=\sin^2 2\theta \sin^2\left(
{1.27 \Delta m^2({\rm eV})^{-2} L({\rm km})^{-1}
\over E({\rm Gev})^{-1}}\right)$$
with the best fit values inside the physical region ($0\le\sin^2 2\theta\le1$)
given as $\sin^2 2\theta=1.0$ and $\Delta m^2=2.2\times10^{-3}$(eV)$^2$.
At the 90\% confidence level,
 $5\times10^{-4}<\Delta m^2< 6\times10^{-3}$(eV)$^2$ and $\sin^2 2\theta>0.82$
({\it ibid.}).

If the oscillation is between $\nu_\tau$ and $\nu_\mu$, no dramatic
changes to GRB models would be called for, as both neutrinos interact
with matter through neutral currents. In the remainder of this letter
I will assume that the detected oscillation is into a sterile flavor
which does not interact with matter. Some features of the data suggest
that this may indeed be the case. Note that a $\mu$  neutrino of energy $E$
would change flavor, with probability one,
from $\mu$ to sterile, every time it traverses a distance
$$L_0= {\pi (E/{\rm GeV}) \over 2.54 \Delta m^2/({\rm eV})^2} {\rm km},
\eqno (1a)$$
and the sterile neutrino would revert back to the $\mu$ flavor after travelling
the same distance again. For a $\sim\,$GeV neutrino, $L_0\sim10^3$km,
but for $E\sim\,$MeV (which may be the more likely value in GRB sources)
the distance is on the order of a kilometer:
$$0.2 (E/{\rm MeV})<(L_0/{\rm km})<2.5(E/{\rm MeV}),
\eqno (1b)$$
where the limits correspond to
the 90\% confidence level limits on $\Delta m^2$. This is a length
scale of some interest in studies of neutron stars (radius $R_*\sim10\,$km)
and solar mass black holes (Schwarzschild radius
$r_s=2.9\,{\rm km}\times M/M_\odot)$, and, by extension, also in theoretical
discussions of GRB models.

\section{Hypernova-type models of GRBs}

There is a class of models in which GRBs are ultimately powered by
neutrino emission in the gravitational collapse of Earth sized (but
massive) objects.  These could be the cores
($M>>M_\odot$) of very massive stars (Woosley 1993, Paczy\'nski 1998),
or those of supernovae type Ic (as possibly in SN 1988bw:
Iwamoto {\it et al.} 1998, Wang \& Wheeler 1998),
an unstable supermassive ($10^5 M_\odot$) quasi-star (Fuller \& Shi 1998),
etc. While a sufficiently high energy input (up to $\sim 10^{54}$erg)
in neutrinos or in Poynting flux can be obtained (refs. cited)
in a natural way, it has never been demonstrated how to avoid the
baryon-loading problem in these cases.

One may expect a significant fraction of a solar mass in baryonic
matter distributed in a ``mantle'' of at least $10^4$km extension in
every direction in these models.  A fraction (up to
a few percent) of the released energy which is sufficiently high
to power the observed GRBs can be
deposited above the mantle, if a comparable energy is
converted into GeV or TeV neutrinos.  By analogy with AGNs, or Galactic
``mini-quasars'' (thought to be stellar-mass binary systems), one would
imagine formation of jets in the system and then shock acceleration of protons,
which could photo-produce pions decaying into neutrinos.

For ordinary quasars, $\mu$-neutrino luminosities as high as
20\% of the quasar luminosity have been predicted (Stecker 1993).
 If a comparable fraction
of the luminosity of the GRB central engine (cloaked in the mantle) were
converted to $\nu_\mu$'s, many of these could penetrate the mantle after
oscillating into the sterile flavor, oscillate back into $\mu$-neutrinos
above the mantle (in agreement with eq. [1])
and then deposit their energy by electron pair creation
and scattering on electrons in a relatively baryon-free region, thus creating
the type of ultra-relativistic outflow discussed in GRB scenarios.

\section{GRB models with solar-mass central engines}

In another class of models, GRBs are powered by a compact object of
about solar mass. This class includes double neutron star systems in
the final stages of binary coalescence, or the interaction of a
neutron star with a black hole in a coalescing binary, neutron stars
with millisecond rotation periods, neutron stars suffering a phase
transition, perhaps to a strange star, and many variants on the above
(Lattimer \& Schramm 1976; Paczy\'nski 1986; Eichler {\it et al.}
1988; Paczy\'nski 1991; Usov 1992; Mochkovitch {\it et al.} 1993; Cheng \& Dai
1996; Klu\'zniak \& Ruderman 1997, 1998; Klu\'zniak \& Lee
1998). Frequently, the discussed configuration is that of a solar-mass
disk orbiting a stellar-mass black hole.  In all these cases, the
typical length scale for the distribution of baryons is a few km, and
up to few times $10^{53}\,$erg of energy can be released. In most (but
not all) cases thermal emission of neutrinos is expected, necessarily
with energies of a few MeV, neutrino emission in such models bears
similarity to that in supernovae. Whether the expected luminosity of
neutrinos is sufficient to power a GRB is a subject of some
controversy, but assuming that it is, the baryon-loading problem is
almost always a serious concern.

Clearly, by eqs. (1) oscillation over a distance comparable to the size
of the system is expected, which would be helpful in circumventing the
baryon-loading problem, as the sterile neutrinos could freely penetrate 
baryonic matter, and even over-take any non-relativistic wind, and deposit
energy in a relatively baryon-free region after oscillating back into
$\mu$-neutrinos.

\section{The geometry of neutrino annihilation}

As the ultra-relativistic blast wave would be powered by
annihilation of $\mu$-neutrinos
into electron positron pairs, it is necessary to discuss the rate of energy
deposition in this process.
The overall efficiency of energy conversion into gamma-rays has to be rather
high in most scenarios, so if sterile (or any) neutrinos are to play a role
in the energy transport, it helps if the geometry of neutrino annihilation
(and simultaneous electron pair creation) is favorable.

In spherical geometry, the luminosity (per unit volume)
of neutrino annihilation drops
like the radius to the eight power, $l_{\nu\bar\nu}(r)\propto r^{-8}$. 
In the discussion above, the
neutrino oscillation was taken to occur over a distance comparable
with the system size, this would correspond to an
efficiency of energy transport by the sterile
neutrinos of $\approx2^{-5}=0.03$ [this is the ratio of
energy deposited in electron positron pairs at $r>2r_0$
to that which would be deposited in the absence of
baryons and neutrino oscillations at $r>r_0$]. However, it is agreed by most
authors that angular momentum in the system would lead to significant
departures from spherical symmetry. Indeed, disk-like or toroidal
geometry seems preferred, and it has been pointed out that
such geometry is more favorable for annihilation when no baryons are present
(M\'esz\'aros \& Rees 1992).
I show below that in disk-like geometry,
when baryons are present through which 
sterile neutrinos transport energy away from the source over a distance
comparable to the oscillation length, the efficiency is a few percent,
i.e. it is comparable to that in the spherical case. A toroidal geometry is
much more favorable and
neutrino annihilation could be further
enhanced by gravitational focusing in that case.

Neglecting gravitational effects (redshift and trajectory bending)
the energy deposition rate, per unit volume, through neutrino annihilation is
(Goodman, Dar \& Nussinov 1987):
$$\dot q(\vec r)={4KG_F^2 \chi(\vec r)\over 3A_0^2c} L_\nu L_{\bar\nu}\left(
{\left<E^2_\nu\right>\over \left<E_\nu\right>}+
{\left<E^2_{\bar\nu}\right>\over \left<E_{\bar\nu}\right>}\right),
\eqno (2)$$
where $L_\nu/A_0$ ($L_{\bar\nu}/A_0$) are the surface emissivities
of neutrinos (antineutrinos),
$K=(1-4\sin^2\theta_W +8\sin^4\theta_W)/(6\pi)=0.027$,
$G^2_F=5.3\times10^{-44}\,{\rm cm}^2{\rm MeV}^{-2}$, and $c$ is the speed
of light.
The expressions in angled brackets are moments of (anti)neutrino energy.
Here, $\chi$ is a
position-dependent geometrical factor.

If the neutrinos and antineutrinos
are emitted isotropically from the surface of a sphere (the same for both)
of radius $r_0$, $L_\nu$ is the neutrino luminosity of that sphere of area
$A_0=4\pi r_0^2$, while
$\chi(\vec r)=\chi_s(x)$, where $\chi_s(x)\equiv (1-x)^4(x^2+4x+5)$,
and $x=\sqrt{1-r_0^2/r^2}$ with $r$ the sperical radial
co-ordinate (Goodman, Dar \& Nussinov 1987). This gives the $r^{-8}$
dependence of luminosity mentioned above: $\chi(\vec r)\approx (5/8)r_0^8/r^8$
for $r^2>>r_0^2$.
Eq. (2) has been derived in the approximation that the netrinos are
sufficiently energetic for threshold effects to be important only over
a negligible solid angle (i.e., $\left<E_\nu\right>>>511\,$keV).

The energy deposition rate at point $P$
on the $z$ axis above some azimuthally
symmetric surface emitting neutrinos isotropically
with surface emissivity 
$L_\nu/A_0$ is given by eq. (2) with
$\chi=\chi_s(x_2)-\chi_s(x_1)$, when the surface is visible from the point $P$
between azimuthal angles $\arccos x_1$ and $\arccos x_2$.
Thus, for an infinite disk $\chi=5$.
If some of the neutrinos are sterile, and do not contribute to the
annihilation rate, there is a slight complication.
As an example consider
a disk isotropically emitting from its surface sterile neutrinos
which oscillate back into the $\mu$ flavor
upon travelling a distance $L_0$, as given by eqs. (1). At point $P$
a height $h$ above the surface, only neutrinos arriving
with azimuthal angle $\theta$ 
satisfying $h/[(2n-1)L_0]<\cos\theta<h/[(2n)L_0]$,
with $n=1$, 2, 3..., are $\mu$-neutrinos and hence contribute to $\dot q$.
Thus, the energy deposition rate is given by eq. (2)
with 
$$\chi=-\chi_s\left({h\over L_0}\right)+\chi_s\left({h\over2 L_0}\right)
 -\chi_s\left({h\over 3L_0}\right) 
+ ... =$$
$$=16 (h/L_0)\ln 2 - {5(\pi h/L_0)^2\over 4}
+ {7(\pi h/L_0)^4\over 144} -{31(\pi h/L_0)^6\over 6\times 7!},$$
for $0<h\le L_0$. E.g., at height $L_0$ above the disk
 the presence of sterile neutrinos reduces the energy deposition rate
by a factor of about two: $\chi=2.503$ for $h=L_0$.

To estimate the efficiency of energy transport through a baryonic mantle
consider a disk, with
no baryons outside, emitting muon neutrinos and antineutrinos and
compute the energy deposition in electron positron pairs up to a
height $L_0$ above the disk, this is obtained by integrating eq. (2)
with $\chi=5$ over this external volume. Now consider an identical
disk emitting sterile neutrinos and antineutrinos at the same
luminosity as before, and assume that the disk is cloaked in baryons,
extending to height $L_0/2$ above the disk.  Taking into account only those
neutrinos which pass through the mantle only in the sterile flavor,
compute the energy
deposition in electron pairs above the baryons, up to the same height
as before, and take the ratio of the two results. The answer is 0.026
and I conclude that the efficiency of energy transport by sterile
neutrinos in disk-like geometry is similar to that in spherical
geometry, on the order of a few percent. This is because neutrinos
travelling sharply away from the normal to the disk surface, oscillate
back into the interacting flavor inside the baryonic mantle, while in
the spherical case the mean collision angle of the neutrinos drops
sharply with distance.

However, for a torus cloaked in baryons to a depth less than $L_0$, the
annihilation rate of $\nu_\mu$'s oscillating out of the sterile phase
is comparable to that for non-oscillating $\mu$-neutrinos in the absence
of a baryonic mantle, i.e. here the efficiency is close to unity.
But it must be noted that in supernova-like emission
the luminosity in muon neutrinos is typically
on the order of only ten percent of the electron luminosity so the overall
efficiency can be no higher than this in such models.

\section{Conclusions}

The Super-Kamiokande collaboration has reported that mu neutrinos
oscillate in vacuum into another non-electronic, flavor, perhaps into
a sterile neutrino. If a significant flux of neutrinos is initially present
in GRB sources (as most models would have it), oscillation of $\nu_\mu$ into
a sterile neutrino would allow the energy to be carried across a region
contaminated with baryons, and would allow its deposition in a relatively
baryon-free region, thus allowing the formation of an ultra-relativistic blast.
The efficiency of this deposition (relative to direct annihilation of
the neutrinos when neither the oscillation nor baryons are present)
is on the order of unity for toroidal geometry of emission, and a few percent
for spherical or disk-like geometry.

In  models involving solar mass compact object (neutron star phase transitions,
mergers etc., Section 4) the oscillation length for the expected
neutrino energy of several MeV matches well the size of the
system. For these models the energy requirements are usually so
extreme that toroidal geometry is required. In
hypernovae models (Section 3), with the characteristic size of the
system on the order of thousands of kilometers, GeV neutrino energies
are required for this mechanism of energy transport to work. In these
models, the available power is usually so large that quasi-spherical
or disk-like geometries with their efficiency of a few percent
suffice.

In this letter I have assumed that it is $\nu_\mu$ that oscillates into
a sterile flavor. The solar neutrino deficit allows (but does not require)
that the electron neutrino oscillates into a sterile flavor. However,
the minimum oscillation length of about 0.2 AU for $\sim\,$MeV neutrinos
required in this case would be too large to be useful in current GRB models.

\acknowledgments
This work supported in part by the KBN through grant 2P03D01311.
I thank the Aspen Center for Physics for hospitality and Mal Ruderman
for comments on the manuscript.

\end{document}